\documentclass[11pt]{article}

\usepackage{amssymb, amsthm}
\usepackage{latexsym}
\usepackage{amsmath}
\usepackage[ansinew]{inputenc}
\usepackage{hyperref}

\numberwithin{equation}{section}

\newtheorem{theo}{Theorem}[section]

\theoremstyle{definition}

\begin{document}
\begin{titlepage}

\date{\today}
\begin{center}
{\bf\Large Quantum mechanics and classical trajectories}

\vskip1.5cm

{\normalsize Christoph N\"olle}

\vskip1cm

{\it Institut f\"ur Theoretische Physik, Leibniz Universit\"at Hannover, \\
 Appelstra{\ss}e 2, 30167 Hannover, Germany}\\
 email: noelle@math.uni-hannover.de

\end{center}

\vskip1.5cm

\begin{abstract}
 The classical limit $\hbar \rightarrow 0$ of quantum mechanics is
 known to be delicate, in particular there seems to be no simple derivation of the classical
 Hamilton equation, starting from Schr\"odinger's equation. In this paper I
 elaborate on an idea of M. Reuter \cite{reuter_QMGaugetheory} to
 represent wave functions by parallel sections of a flat vector
 bundle over phase space, using the connection of Fedosov's
 construction of deformation quantization. This generalizes the
 ordinary Schr\"odinger representation, and allows naturally for a
 description of quantum states in terms of a curve plus a wave
 function. Hamilton's equation arises in this context as a condition
 on the curve, ensuring the dynamics to split into a classical and a
 quantum part.
\end{abstract}
\end{titlepage}
\setcounter{tocdepth}{1}

 \tableofcontents

\section{Introduction}
 
 In the usual formalism of quantum mechanics, where pure states are time-dependent elements of a Hilbert space on which
 observables act as linear operators, the dynamics is governed by Schr\"odinger's equation, whereas in the classical
regime $\hbar \rightarrow 0$ pure states are represented by trajectories in phase space obeying Hamilton's equation and
observables are functions on phase space. In the classical limit $\hbar \rightarrow 0$ one should therefore be able to
associate a curve to a wave function, and to prove that it satisfies Hamilton's equation if the wave function solves
Schr\"odinger's equation. 

\bigskip
 
 This problem turns out to be surprisingly hard, and it has found its most satisfying solution in a somewhat
alternative formulation of quantum mechanics, in the work of Groenewold and Moyal
\cite{Groenewold_QM,
moyal:original}. They introduced the so-called
star product, which allows to treat quantum observables as functions
on phase space, like in the classical regime. Only the algebra structure is
modified, reflecting the non-commutativity of operators. States can then
be
defined very formally as functionals on phase space, both classically and quantized, and the time-dependence is shifted
to the observables, as in the Heisenberg picture. The evolution equation for an observable $f$ reads
 \begin{equation}
   \frac d{dt} f = \frac i\hbar \big( H*_\hbar f-f*_\hbar H\big),
 \end{equation} 
 where $H$ is the Hamilton function of the system, and $*_\hbar $ the Groenewold-Moyal product for
phase space $\mathbb R^{2n}$, or more generally any star-product on a symplectic manifold. In the classical limit $\hbar
\rightarrow 0$ the product $*_\hbar $ is required to satisfy $ (f*_\hbar g - g*_\hbar f ) \rightarrow i\hbar 
\{f,g\}$ for any two observables $f,g$, where on the right hand side we have the Poisson bracket of $f$ and $g$. Thus in
the classical limit we obtain 
 \begin{equation}
   \frac d{dt} f = \{ f,H\},
 \end{equation} 
 which is equivalent to Hamilton's equation in the usual picture of time-dependent states and stationary observables.
 Besides solving the classical limit problem, the star-product approach to quantization has proved very useful and led
to the concept of deformation quantization, culminating in the work of Fedosov \cite{fedosov:deformationart} and
Kontsevich \cite{kontsevich_deformPoisson}.             
               
 \bigskip

 In the present note I propose a reformulation of ordinary quantum mechanics, using ideas from
deformation quantization as well, that allows us to derive Hamilton's equation directly from the
Schr\"odinger equation.
 Instead of working with a deformed product we use Fedosov's method to represent observables as
parallel sections of a flat vector bundle, and consider a particular representation of the resulting algebra, following
ideas of Reuter \cite{reuter_QMGaugetheory}. In the end our method looks very similar to the textbook formulation of
quantum mechanics, and can be formulated without recourse to deformation quantization.

 \bigskip

 We will treat pure quantum states as
equivalence classes $[c,\psi]$, where $c: \mathbb
R\rightarrow \mathbb R^{2n}$ is a curve in phase space and $\psi \in L^ 2(\mathbb R^n)$ is an ordinary wave function.
The
interpretation is that to
every point $\xi$ in our phase space is attached a Hilbert space $\mathcal H_\xi$, all of them isomorphic to $\mathcal
H_0=L^2(\mathbb R^n)$, and that $\psi(t)$ takes values in $\mathcal H_{c(t)}$. In other words we have a trivial Hilbert
bundle $\mathcal
H\rightarrow \mathbb R^{2n}$ over phase space, and $\psi$ is a section of the pullback bundle $c^*\mathcal H$.
 The equivalence relation is defined in terms of a parallel transport operator $U$ which is used to identify the
different Hilbert spaces: for all points $\xi,\sigma\in \mathbb R^{2n}$ we have an isomorphism 
 \begin{equation}
   U(\xi,\sigma) : \mathcal H_\xi \rightarrow \mathcal H_\sigma,
 \end{equation} 
 and we identify two pairs $(c,\psi)$ and $(c',\psi')$ if and only if the relation $\psi'(t) = U(c(t),c'(t) ) \psi(t)$
holds for all $t$. 

\bigskip
 
 The question remains how to define the parallel transport $U$. As observed by Reuter in
\cite{reuter_QMGaugetheory} Fedosov's construction of deformation quantization
yields a connection on our Hilbert bundle $\mathcal H$, and this defines the parallel transport, up to a minor
modification which is necessary because Fedosov's connection is not flat, and therefore its parallel transport is
path-dependent. This will be fixed by tensoring $\mathcal H$ with a line-bundle $B$ with connection, in fact the
prequantum line-bundle from geometric quantization. The resulting bundle has indeed a flat connection
\cite{Noelle_GeoDef2}.
 Instead of using pairs $[c,\psi]$ to describe a pure state we
can then equivalently use a time-dependent parallel section $\Psi$ of $\mathcal H\otimes B$. Its value at $\xi\in
\mathbb
R^{2n}$ is obtained by parallel-transport of $\psi(t)$:
 \begin{equation}
   \Psi_\xi (t) = U\big(c(t),\xi\big) \psi(t) \qquad \in (\mathcal H \otimes B)_\xi.
 \end{equation} 

Similarly, the operator $\rho(f)$ corresponding to a classical observable $f$ can be represented by a parallel section
w.r.t. the
induced connection on the endomorphism bundle. This is the main idea of Fedosov's
construction.

\bigskip

 Due to the equivalence relation the choice of the curve $c$ to represent a state $[c,\psi]$ is completely
arbitrary: for every pure state $\eta$ and every curve $c$ we can find a wave function $\psi$ such that $\eta =
[c,\psi]$. In particular,
choosing $c$ to remain in the origin, $c \equiv  0$, we recover the textbook formulation of quantum mechanics, but for
other trajectories we get different (equivalent) representations.
 The dynamics in this picture is governed by what we call the Schr\"odinger equation over $c$. As usual this is a
differential equation for $\psi$, whose explicit form depends on the choice of $c$, however. If $c$
is constant it
looks similar to the usual Schr\"odinger equation, but for non-constant $c$ an important modification occurs. 
 
\bigskip

 As any curve can be used to represent a given quantum state, in combination with an appropriate wave function,
we can ask whether there is a suitable notion of preferred trajectories. Taking a closer look at expectation values of
observables we find that they naturally split into a classical and a quantum contribution, where the classical part is
uniquely determined by the trajectory $c$ alone. Apparently this splitting then depends on the choice of representative
for the equivalence class $[c,\psi]$. We will define the preferred trajectories as those where the quantum contribution
to expectation values remains small (i.e. of order $\hbar$) under time evolution. The main result of the paper is that
these preferred trajectories are exactly the solutions of Hamilton's equation. 

\bigskip

 This can be understood as a derivation of Hamilton's equation from the Schr\"odinger equation. Starting with a quantum
state $[c,\psi]$ solving Schr\"odinger's equation, the measurement of any observable $f$ in this state in the classical
limit $\hbar \rightarrow 0$ gives the value $f(c(t))$ if $c$ solves Hamilton's equation with appropriate
boundary condition, and the disappearance of the wave function in the classical limit is not at all mysterious any
more. It is a consequence of a good choice of representative for every state.

\bigskip

 The derivation of this result requires some assumptions on the
 initial wave function $\psi(t_0)$, which are summarized in theorem \ref{theo}. Basically one has to make sure that
 the wave function does not
 spread over a macroscopic area (neither in position space, nor in
 momentum space when Fourier transformed). This could have been expected, because even classically the center of
mass motion in this case would not be described by a trajectory solving Hamilton's equation, but by some phase
space density 
 obeying Liouville's equation. 
 Finally, the generalization to quantizable symplectic
 manifolds is indicated in section \ref{sec:CPS}.

\section{Quantum mechanics over phase space}

\paragraph{Deformation quantization.}
 We consider a trivial Hilbert bundle $\mathcal H\rightarrow \mathbb R^{2n}$ over phase space, i.e. a collection of
Hilbert spaces $\mathcal H_\xi\simeq L^2(\mathbb R^n)$ for every point $\xi \in \mathbb R^{2n}$.
 They come equipped with an action of the Weyl algebra, generated by $\hat q^j$ and
 $\hat p_k$ (where $j,k=1,\dots,n$):
  $$ \hat q^j\psi(x)= x^j\psi(x),\qquad
 \hat p_k\psi(x) = \frac \hbar i\frac \partial{\partial
 x^k}\psi(x)$$
 for $\psi \in L^2(\mathbb R^n)$ and $x\in \mathbb R^n$.
 We will denote these generators collectively by $y^\mu$ ($\mu=1,\dots,2n$), they satisfy the canonical commutation
relations $[y^\mu,y^\nu]=i\hbar\omega^{\mu\nu}$, where $\omega=dp_j\wedge dq^j$ is the standard symplectic form on
$\mathbb R^{2n}$ with linear coordinates $q^j,p_k$ ($j,k=1,\dots n$).  
  States $\Psi$ will be sections of the Hilbert bundle, so that they depend both on the phase space point $\xi$
as well as an auxiliary variably $x\in \mathbb R^n$.
 Since we introduced additional degrees of freedom this way as compared to the usual formulation of quantum mechanics,
we need to impose a further constraint on physical states, and we require them to be parallel:
 \begin{equation}
  D\Psi=0,
 \end{equation} 
 where
  \begin{equation}\label{FedCon}
    D = d-\frac i\hbar \Big[\theta + \omega_{ab}y^a d\xi ^b\Big]
  \end{equation}
 is a connection on our trivial Hilbert bundle, whith $d=dq^j\frac {\partial}{\partial q^j} + dp_j \frac
{\partial}{\partial
 p_j}$ being the exterior derivative, and $\theta$ any 1-form on $\mathbb R^{2n}$
 satisfying $d\theta=\omega$. The results will be independent of the
 precise form of $\theta$, but for the example below a convenient
 choice turns out to be $\theta= \frac 12(p_a dq^q - q^a dp_a)$, which we
 therefore adopt. The solutions of $D\Psi=0$ have the form
 \begin{equation}
  \Psi_{(q,p)}(x) = \chi(q+x) e^{-\frac i\hbar p(x+\frac q2)},
 \end{equation}
 where $\chi$ is any square-integrable function on $\mathbb R^n$, and for $p,x\in \mathbb R^n$ the expression $px$
denotes the standard
inner product $p_jx^j$.
 The operator $\rho(f)$ corresponding to a function $f$, satisyfing
$[D,\rho(f)]=0$ and acting on
$\mathcal H_{\xi}$, is
 \begin{align}\label{OperatorExp}
  \rho( f)_{\xi}& = \sum_{k=0}^\infty \frac 1{k!} \partial_{\mu_1}
  \dots \partial_{\mu_k} f(\xi) y^{\mu_1}\dots y^{\mu_k}  \\
    &=f(\xi) +
  \partial_\mu f(\xi) y^\mu + \frac 12 \partial_\mu \partial_\nu f
  (\xi) y^\mu y^\nu+\dots,\nonumber
 \end{align}
 where $y^{\mu_1}\dots y^{\mu_k}$ denotes the symmetrized product of
 the operators $y^{\mu_1},\dots,y^{\mu_k}$ \cite{fedosov:deformationart}. If we choose the base point equal to the
origin, i.e. $\xi=0$, then \eqref{OperatorExp} reproduces the standard quantization prescription in Weyl ordering.
 Finally, the parallel transport operator 
\begin{equation}
 U:=
 U\big((q_0,p_0),(q,p)\big): \mathcal H_{(q_0,p_0)} \rightarrow \mathcal H_{(q,p)},
\end{equation} 
 defined by $\Psi_{(q,p)} = U
 \Psi_{(q_0,p_0)}$ for every physical state $\Psi$, is a Weyl operator
 \begin{equation}
  U = \exp \Big[ \frac i\hbar \Big((p_0-p)\hat q + (q-q_0)\hat p
  +\textstyle{\frac 12} (qp_0-p q_0)\Big)\Big].
 \end{equation}
The important property we need is that $U$ satisfies the
 parallel transport equation
  \begin{equation}
   \partial_t U\big(\xi,c(t)\big) = -A_{c(t)}\big(\dot c(t)\big)\,
   U\big(\xi,c(t)\big),
  \end{equation}
 where $\xi\in \mathbb R^{2n}$, $c$ is any curve starting in $\xi$, and $A$ is the connection 1-form of $D$, i.e.
  \begin{equation}
    A = -\frac i\hbar \Big[\theta + \omega_{ab}y^a d\xi^b\Big].
  \end{equation}

\paragraph{Dynamics over a fixed base point.}
 Let $H$ be a classical Hamilton function and $\psi \in \mathcal H_\xi$ a wave function over $\xi\in \mathbb R^{2n}$.
 Then the Schr\"odinger equation for $\psi$ is
 \begin{equation}\label{SchroedingerBaseFix}
  i\hbar\,\partial_t \psi = \rho( H)_\xi \, \psi,
 \end{equation}
  and the expectation value of an observable $f\in C^\infty(\mathbb
 R^{2n})$ is given by (we assume $\psi$ to be normalized)
 \begin{equation}\label{ExpValueBaseFix}
  \langle f\rangle _\psi = \langle \psi, \rho( f)_\xi \psi\rangle
  =f(\xi) +\partial_\mu f(\xi) \langle \psi,y^\mu \psi\rangle +\frac
  12 \partial_\mu\partial_\nu f(\xi)\langle \psi,y^\mu y^\nu
  \psi\rangle +\dots
 \end{equation}
 The first term gives a classical contribution, i.e. just the value
 of the function $f$ at $\xi$, the next terms are quantum
 corrections. A natural expectation is that the latter ones should be
 small compared to the classical contribution, but this
 cannot be true in general, as the classical part is determined by the completely arbitrary choice of base point and
remains constant in
 time. To make this more precise, we will consider $\hbar$ as a
 formal variable from now on (partly motivated by Fedosov's
 strategy in \cite{fedosov:deformationart}). Wave functions will have to be considered as formal Laurent series in 
some power of $\hbar$ then, whereas expectation
 values of $\hbar$-independent observables should become power series in $\hbar^{1/2}$, and we say a term is small
 or microscopic if it is of order $\hbar^{1/2}$ at least, and large or
 macroscopic, if it is of order $\hbar ^0=1$. 

\bigskip

 The relation $[y^\mu,y^\nu]=i\hbar \omega^{\mu\nu}$ suggests to
 assign the $\hbar$-degree $1/2$ to each of the $y^\mu$, and to
 expect terms $\langle \psi,y^{\mu_1}\dots
 y^{\mu_k}\psi\rangle$ to be of $\mathcal O(\hbar^{k/2})$. This
 $\hbar$-grading has turned out to be a very powerful tool in the
 construction of quantum deformations of general curved phase spaces
 \cite{fedosov:deformationart}, and it is perhaps surprising to find it not being
 respected by taking expectation values.
 To see explicitly why this expectation fails here, let us consider
 the time evolution of expectation values of the simple coordinate functions $q^j$ and $p_k$. Let
 $\xi = (a,b)$, where $a \in \mathbb R^n$ are the $q$-components,
 and $b$ the $p$-components, then the splitting into classical and quantum contributions is
  \begin{equation}\label{SimpleExpVals}
   \langle q^j\rangle _\psi= a^j + \langle \psi,\hat q^j
   \psi\rangle,\qquad  \langle p_k\rangle _\psi = b_k + \langle \psi,\hat p_k
   \psi\rangle.
  \end{equation}
 The Schr\"odinger equation for $\psi$ implies
  \begin{align}
   \partial _t\big\langle \psi ,y^\mu \psi\big\rangle &= \frac i\hbar \big\langle \psi
  ,[\rho(H),y^\mu]\psi\big\rangle \nonumber\\
    &=  \frac i\hbar\Big[\partial _\alpha H \big\langle \psi, [y^\alpha,y^\mu]\psi\big\rangle + \frac 12 \partial_\alpha
    \partial_\beta H \big\langle \psi, [y^\alpha y^\beta,y^\mu]\psi\big\rangle +\ \dots,\Big]
    \nonumber\\
    &=\underbrace{\omega^{\mu\alpha}\partial_\alpha H}
    +\underbrace{\omega^{\mu\alpha}\partial_\alpha\partial_\beta H \langle
    \psi|y^\beta \psi\rangle }+\ \underbrace\dots \\
     &\qquad\quad \hbar^0 \qquad\qquad \quad\ \hbar^{1/2} \qquad\qquad \quad\mathcal
     O(\hbar^{1})\nonumber
  \end{align}
 where the indicated $\hbar$-degrees are the ones one would naively
 expect, according to the filtration explained above.
 The first term on the right hand side is responsible for the fact that
 under time evolution the quantum part $\langle \psi,\hat
 q^j\psi\rangle$ assumes macroscopic values, even if it was
 of order $\hbar^{1/2}$ initially. Then by \eqref{ExpValueBaseFix} this is also true for
 the quantum corrections of other observables. If the operator
 $\rho(H)=H+\partial_\mu H y^\mu + \frac 12 \partial_\mu\partial_\nu Hy^\mu y^\nu+\dots$
 had no terms linear in $y^\mu$, the $\hbar$-filtration would be preserved under time-evolution.

\paragraph{Dynamics over a curve.}
 Let us then consider the Schr\"odinger equation over a curve. Suppose that $\psi \in \mathcal H_\xi$ is a wave
function solving the Schr\"odinger equation \eqref{SchroedingerBaseFix}, and $c$ a curve starting in $\xi$. Then the
wave function $\phi \in \Gamma(c^* \mathcal H)$, defined by 
  $$\phi(t) := U\big(\xi,c(t)\big) \psi(t) \ \in \mathcal
  H_{c(t)}$$
 satisfies the Schr\"odinger equation along $c$:
 \begin{align}\label{SchroedingerOverCurve}
  i\hbar\, \partial_t \phi &= i\hbar (\partial_t U) \psi +
  i\hbar U\partial_t  \psi \nonumber \\
   &= -i\hbar A(\dot c)U\psi  + U\rho( H)_\xi\psi \\
   &= \big(\rho( H)_{c(t)} - i\hbar A_{c(t)}(\dot
   c(t))\big)\phi(t),\nonumber
 \end{align}
 where we used that $\rho(H)_{c(t)} = U\rho(H)_\xi U^{-1}$. Inserting the explicit expressions for $\rho(H)$ and $A$ we
find that 
 \begin{equation}
   i\hbar\, \partial_t \phi = \Big(H - \theta (\dot c) + \big(\partial_\mu H - \omega_{\mu\nu}\dot c^\nu \big) y^\mu +
\frac 12 \partial_\mu\partial_\nu H y^\mu y^\nu +\dots\Big)\Big|_{c(t)} \phi(t) 
 \end{equation} 
 Only the terms of $\hbar$-degree 0 and 1/2 have changed compared to the usual Schr\"odinger equation
\eqref{SchroedingerBaseFix}, and the degree 1/2 term vanishes if and only if $c$ satisfies Hamilton's equation
  \begin{equation}\label{HamiltonEq}
   \partial_t c^\mu(t)  = \omega^{\mu\nu} \partial_\nu H(c(t)).
  \end{equation}
In this case we get
 \begin{equation}\label{SchroedingerCurveExplicit}
  i\hbar\, \partial_t \phi = \Big(H-\theta_\mu
  \omega^{\mu\nu}\partial_\nu H + \sum_{k=2}^\infty \frac 1{k!}
  \partial_{\mu_1}\dots \partial_{\mu_k} H y^{\mu_1}\dots
  y^{\mu_k}\Big)\Big|_{c(t)} \phi,
 \end{equation}
and the $\hbar$-filtration is preserved under taking commutators with the 
  modified Hamilton operator on the right hand side. The time evolution of the
 lowest quantum correction $\langle\phi,y^\mu \phi\rangle$ becomes
  \begin{equation}
   \partial_t \langle \phi,y^\mu\phi\rangle = \omega^{\mu\alpha}\partial_\alpha\partial_\beta H \langle
    \phi,y^\beta \phi\rangle +  \mathcal O(\hbar^1),
  \end{equation}
 and similarly for the expectation values of higher powers of the
 canonical operators.
 Therefore, if initially the $\hbar$-filtration is respected by the expectation values, so that
 $\langle \phi(t_0), y^{\mu_1}\dots y^{\mu_k} \phi(t_0)\rangle$ is
 of order $ \hbar^{k/2}$, then this filtration will be preserved under time-evolution.
 Accordingly, the quantum
 corrections to the classical result $f(c(t))$ of the expectation value
 \begin{equation}
  \langle f\rangle _\phi = f(c(t)) +\partial_\mu f(c(t)) \big\langle
  \phi,y^\mu \phi\big\rangle + \frac 12\partial_\mu\partial_\nu f(c(t))
  \big\langle \phi ,y^\mu y^\nu \phi\big\rangle +\dots
 \end{equation}
 remain of order $\hbar^{1/2}$ and vanish in the
 classical limit $\hbar\rightarrow 0$. In other words, we
 have deduced Hamilton's equation from the Schr\"odinger equation by
 taking the classical limit. To collect the assumptions, we
 formulate the result as a theorem:
\begin{theo}\label{theo}
 Let $\phi(t) \in L^2(\mathbb R^n)$ satisfy the Schr\"odinger
 equation \eqref{SchroedingerOverCurve} over a curve $c$ solving Hamilton's
 equation \eqref{HamiltonEq}. Assume that at some initial time $t_0$
 the expectation values of the canonical operators $y^\mu$ satisfy
  \begin{equation}
   \big\langle \phi (t_0),y^{\mu_1}\dots y^{\mu_k} \phi(t_0)\big\rangle
    = \mathcal O(\hbar^{k/2})
  \end{equation} 
 for all $k \geq 1$ and $\mu_j=1,\dots,2n$. Then the classical limit of
 the expectation value $\langle f\rangle _{\phi(t)}$, for an observable $f\in C^\infty(M)$, is
 $$  \lim_{\hbar\rightarrow 0} \langle f\rangle _{\phi(t)}=
 f(c(t)).$$
\end{theo}
 The limit $\hbar\rightarrow 0$ means of course to discard all terms of positive
 $\hbar$-degree. The assumptions on the expectation values are
 satisfied e.g. by the eigenfunctions of the harmonic oscillator,
 and also by coherent states. For the latter ones the theorem has
 been proven in \cite{Hepp74} already. 

\bigskip

 What is the physical meaning of these conditions? If
 e.g. $\langle \hat q^2 \rangle$ is of order $\hbar^0= 1$, then the wave
 function spreads over a macroscopic area, and even classically
 one does not expect the center of mass motion to coincide with that
 of a localized point particle. It is then tempting to ask whether
 there is a density $\rho$ on phase space which under time
 evolution w.r.t. the Liouville equation gives rise to the same
 expectation values as $\phi$, up to quantum corrections, even
 without the localization condition. Here we
 shall not pursue this question further, but content us with the
 explanation of the assumption in the theorem.

\section{Example: harmonic oscillator}
 Consider the Hamiltonian
  \begin{equation}
    H(q,p) = \frac 12\big(p^2+q^2\big).
  \end{equation}
 on a two-dimensional phase space.
 Its corresponding quantum operator is
 \begin{equation}
   \rho(H)_{(q,p)} = \frac 12\big(q^2+p^2\big) + q\hat q +p \hat p + \frac
   12 \big(\hat q^2+\hat p^2\big).
 \end{equation}
 Classical solutions $c: \mathbb R\rightarrow \mathbb R^2$ of Hamilton's equation
  $$ \partial_t c = \left(
                      \begin{array}{cc}
                        0 & 1 \\
                        -1 & 0 \\
                      \end{array}
                    \right) c
  $$
 are of the form
 \begin{equation}\label{OscTrajectory}
  c(t) = \exp\Big\{ \left(
                      \begin{array}{cc}
                        0 & 1 \\
                        -1 & 0 \\
                      \end{array}
                    \right)t
   \Big\} c(0) = \left(\begin{array}{cc}
                   \cos\ t & \sin\ t \\
                   -\sin\ t & \cos\ t
                 \end{array} \right)c(0).
 \end{equation}
 Then the Schr\"odinger equation along $c$, equation
 \eqref{SchroedingerCurveExplicit} with $\theta =\frac 12( pdq-qdp)$,
 reads
 \begin{equation}
  i\hbar \partial_t \phi = \frac 12\big(\hat p^2+\hat q^2\big)\phi,
 \end{equation}
 i.e. the ordinary oscillator equation, although we are working
 over the curve $c$ now. So if $\phi$ initially is an eigenstate $|n\rangle$ of $\frac
 12(\hat p^2 +\hat q^2)$, it remains so and describes quantum fluctuations around the motion of a particle
 moving along the classical trajectory \eqref{OscTrajectory}. As $H$
 contains no monomials of degree higher than 2 in $q$ and $p$, and $\langle n|y^\mu n\rangle =0$,
 equation \eqref{SimpleExpVals} implies that
  \begin{equation}
        \langle q \rangle_\phi = c^q(t) = \cos(t) c^q(0) + \sin(t)
  c^p(0),
  \end{equation}
 and there are no quantum corrections at all to the center of mass
 motion. What do these states look like in the standard Hilbert
 space $\mathcal H_0$? Let's take $\phi(0)$ as the ground state $|0\rangle$, then
 $\phi(t) =e^{-\frac {it}2}|0\rangle$, and
 the corresponding wave function of textbook quantum mechanics, $\psi\in\mathcal H_0$, is given by
 \begin{equation}
  \psi(t)=U\big(c(t),0\big)\phi(t) = \exp\Big[\frac i\hbar
  \Big(c^p(t)\hat q -c^q(t)\hat p\Big)\Big] \phi(t).
 \end{equation}
 In terms of the creation and annihilation operators
  $$ \hat a = \sqrt{\frac 1{2\hbar}}\big(\hat q+i\hat p\big) ,\qquad
  \hat a^\dagger = \sqrt{\frac 1{2\hbar}}\big(\hat q-i\hat p\big)$$
 this reads
 \begin{equation}
  \psi(t) = e^{-\frac {it}2}\exp\Big(ze^{-it}\hat a^\dagger -
  \overline z e^{it}\hat a\Big)|0\rangle,
 \end{equation}
 with $z= c^q(0)+ ic^p(0) $. These are the 
 'coherent states', which resemble classical solutions as closely as possible.

\section{Curved phase spaces}\label{sec:CPS}
 Theorem \ref{theo} generalizes to the case of an arbitrary
 symplectic phase space $(M,\omega)$ satisfying the usual quantization conditions
 that $\frac{ [\omega]}{2\pi\hbar}\in H^2(M,\mathbb Z)$ and $c_1(M)$
 is even. In this case there is again a Hilbert bundle $\mathcal H\rightarrow
 M$, the bundle of 'metaplectic spinors', and one can find a so
 called prequantum line bundle $B\rightarrow M$. On $\mathcal H$
 one has Fedosov's connection at one's disposal, of curvature $\frac i\hbar
 \omega$, and $B$ carries a connection of curvature $-\frac i\hbar
 \omega$. The tensor product $\mathcal H\otimes B$ thus has a flat
 connection, and the wave functions are parallel sections of
 $\mathcal H\otimes B$. Furthermore there is an action of the observables $C^\infty(M)$
 on the space of wave functions. At least on cotangent bundles one can prove
 that this representation of quantum mechanics is equivalent to (an
 extension of) the well-known geometric quantization of $M$
 \cite{Noelle_GeoDef2}. It seems however that the representation
 considered here
 is of little use in practice, as neither the wave functions
 nor the observables can be calculated exactly.\\
 The construction depends on the choice of a torsion-free symplectic
 connection $\nabla = d+\Gamma$ on $M$, but different choices lead
 to equivalent quantizations. The explicit form of the operator $\rho(f)$ corresponding to $f\in
 C^\infty(M)$ is
 \begin{equation}
  \rho(f) = f + \partial_\mu f y^\mu + \frac
  12\big(\partial_\mu\partial_\nu - \Gamma^\kappa _{\mu\nu}
  \partial_\kappa \big)f y^\mu y^\nu + \mathcal O(\hbar^{3/2}),
 \end{equation}
 and the connection form of the tensor product connection on $\mathcal H\otimes B$ is
  \begin{equation}
        A= -\frac i\hbar \big(\theta + \omega_{\mu\nu}y^\mu dx^\nu\big)
        -\frac i{2\hbar}\Gamma_{\kappa\mu\nu}
        y^\kappa y^\mu dx ^\nu - \frac
        i{8\hbar}R_{\kappa\lambda\mu\nu} y^\kappa y^\lambda y^\mu
        dx^\nu  + \mathcal O(\hbar ^{1}),
  \end{equation}
 where $R$ is the curvature of $\nabla$; for details consult \cite{fedosov:deformationart}. To find the explicit form of
the Schr\"odinger equation \eqref{SchroedingerOverCurve} we need to evaluate $A$ on the velocity
 vector of a curve satisfying Hamilton's equation
 $\partial_t c^\mu = \omega^{\mu\nu}\partial_\nu H$. This gives
 \begin{equation}
  -i\hbar A(\dot c) = -\theta_\mu\omega^{\mu\nu}\partial_\nu H -
  \partial_\mu H y^\mu +\frac 12
  \Gamma^\kappa_{\mu\nu}\partial_\kappa H y^\mu y^\nu - \frac 18
  R_{\kappa\lambda \mu\nu}\omega^{\nu\alpha}\partial_\alpha H
  y^\kappa y^\lambda y^\mu + \mathcal O(\hbar^2).
 \end{equation}
 The Schr\"odinger equation along $c$, $i\hbar \partial_t \phi =
 (\rho(H) - i\hbar A(\dot c))\phi$, becomes
 \begin{equation}
  i\hbar\partial_t \phi  =
  \Big(H-\theta_\mu\omega^{\mu\nu}\partial_\nu H +\frac 12
  \partial_\mu\partial_\nu H y^\mu y^\nu +\mathcal
  O(\hbar^{3/2})\Big) \phi,
 \end{equation}
 looking exactly as in the flat case up this order in $\hbar$.
 Therefore the $\hbar$-degree of the expectation values $\langle
 y^{\mu_1}\dots y^{\mu_k}\rangle $ is again not lowered under time evolution
 along $c$, and theorem \ref{theo} remains valid.

\bigskip

 One should note however that the quantization method explained in this section is not rigorous. The connection
form
on $\mathcal H\otimes B$ is an infinte power series in $\hbar$ and the canonical operators $y^\mu$, and there seems to
be little hope to prove convergence in any sensible way. In certain cases it is possible however to prove that the
induced formal star-product on a subalgebra of the observables involves only finitely many $\hbar$-powers
\cite{BordemannNW-StarprodKot, Tamarkin:symmKaehler}, and that it has a well-defined representation which is at least
formally equivalent to the one indicated here \cite{Noelle_GeoDef2}. 

\bibliographystyle{Lit}
\bibliography{literatur}

\end{document}